\documentclass[%
 reprint,
amsmath,amssymb,
aip,
apl,
nofootinbib
]{revtex4-1}
\usepackage{dcolumn}

\usepackage{amsmath}
\usepackage{lipsum}
\usepackage{filecontents}
\usepackage[usenames,dvipsnames,svgnames,table]{xcolor}
\usepackage{graphicx}
\usepackage[caption=false]{subfig}
\usepackage{comment}
\usepackage[utf8]{inputenc}
\usepackage[T1]{fontenc}
\usepackage{chngcntr}
\usepackage{appendix}
\usepackage{ifpdf}
\ifpdf
  \DeclareGraphicsExtensions{.pdf,.png,.jpg}
\else
  \DeclareGraphicsExtensions{.eps}
\fi
\usepackage[scale=0.86]{geometry}
\usepackage{color,soul}
\usepackage[framemethod=tikz]{mdframed}
\usepackage{import}
\usepackage{float}
\usepackage{layout}
\usepackage{bm}
\usepackage{lipsum}
\usepackage{fancyhdr}

\fancypagestyle{Text}{
\fancyhf{}
\lhead{Aryan Mehboudi and Junghoon Yeom}
\chead{}
\rhead{\today}
\rfoot{{\normalsize Page \thepage}}

}

\fancypagestyle{Appendices}{
\fancyhf{}
\lhead{A. Mehboudi and J. Yeom}
\chead{\Large {\bf {Supplemental Material}}}
\rhead{\today}
\rfoot{{\normalsize Page \thepage}}

\pagenumbering{arabic}
\setcounter{page}{1}
\setcounter{equation}{0}
\setcounter{figure}{0}
\renewcommand\thepage{A.\arabic{page}}

}

\fancypagestyle{Appendix_A}{
\fancyhf{}
\lhead{A. Mehboudi and J. Yeom}
\chead{\Large {\bf {Supplemental Material}}}
\rhead{\today}
\rfoot{{\normalsize Page \thepage}}

\pagenumbering{arabic}
\setcounter{page}{1}
\setcounter{equation}{0}
\setcounter{figure}{0}
\renewcommand\thepage{A.\arabic{page}}

}

\fancypagestyle{Appendix_B}{
\fancyhf{}
\lhead{A. Mehboudi and J. Yeom}
\chead{\Large {\bf {Supplemental Material}}}
\rhead{\today}
\rfoot{{\normalsize Page \thepage}}

\pagenumbering{arabic}
\setcounter{page}{1}
\setcounter{equation}{0}
\setcounter{figure}{0}
\renewcommand\thepage{B.\arabic{page}}

}

\usepackage{hyperref}
\usepackage{cleveref}

\graphicspath{{src/figs/}}

\hypersetup{
    colorlinks,
    citecolor=red,
    filecolor=green,
    linkcolor=purple,
		urlcolor=yellow
}

\newcommand\myshade{85}
\colorlet{mylinkcolor}{violet}
\colorlet{mycitecolor}{YellowOrange}
\colorlet{myurlcolor}{Aquamarine}

\hypersetup{
  linkcolor  = mylinkcolor!\myshade!black,
  citecolor  = mycitecolor!\myshade!black,
  urlcolor   = myurlcolor!\myshade!black,
  colorlinks = true,
}
\AtBeginDocument{%
  \hypersetup{
		citecolor=Red,
		linkcolor=violet,   
		urlcolor=blue
		}}

\begin{document}

\title{A passive Stokes flow rectifier for Newtonian fluids}

\author{Aryan Mehboudi}
\email[]{mehboudi@egr.msu.edu}

\affiliation{Mechanical Engineering Department, Michigan State University, East Lansing, Michigan, US}

\author{Junghoon Yeom}
\email[]{jyeom@egr.msu.edu}

\affiliation{Mechanical Engineering Department, Michigan State University, East Lansing, Michigan, US}

\date{\today}

\begin{abstract}
Non-linear effects of the Navier-Stokes equations disappear under the Stokes regime of 
Newtonian fluid flows 
disallowing the fluid flow rectification.
Here we show mathematically and experimentally that passive flow rectification of Newtonian fluids is obtainable under the Stokes regime of 
both compressible and incompressible flows
by introducing nonlinearity into the otherwise linear Stokes equations. 
Asymmetric flow resistances arise in shallow nozzle/diffuser microchannels with deformable ceiling, in which the fluid flow is governed by a non-linear coupled fluid-solid mechanics equation.
Fluid flow rectification has been demonstrated for low-Reynolds-number flows (Re $\sim O(10^{-3})$~\textemdash~$O(10^{0})$) of common Newtonian fluids such as air, water, and alcohol.
This mechanism can pave the way for regulating the low-Reynolds-number fluid flows with potential applications in precise low-flow-rate micropumps, drug delivery systems, etc.
\end{abstract}
\maketitle

\newpage

Directional dependence of the hydrodynamic resistance in asymmetrically-shaped microchannels results in the fluid flow rectification, which is important for different applications such as integrated microfluidic devices \cite{leslie_frequency-specific_2009,mosadegh_integrated_2010,weaver_static_2010,mosadegh_next-generation_2011}, micropumps \cite{stemme_valveless_1993,olsson_micromachined_1997,amirouche_current_2009} and drug delivery \cite{nisar_mems-based_2008,eddington_flow_2004,tandon_microfabricated_2016}.
In particular, the rigid nozzle/diffuser microchannels are widely used to rectify the fluid flow at sufficiently high Reynolds numbers (Re $\gg 1$) \cite{stemme_valveless_1993,olsson_micromachined_1997}, wherein the underlying mechanism is rooted in the directional dependence of the hydrodynamic resistance due to the non-linear nature of the Navier-Stokes equations.
Under the Stokes flow regime (Re $\ll 1$), however, the inertial terms forming the non-linear behavior of the governing equations diminish, causing the Newtonian fluid flow through a rigid nozzle/diffuser microchannel to exhibit a direction-independent hydrodynamic resistance disallowing the fluid flow rectification.

To realize a fluidic rectifier that operates under the Stokes flow regime, 
researchers have resorted to another source of nonlinearity, that is, the use of non-Newtonian working fluids with nonlinear rheological characteristics \cite{groisman_microfluidic_2003,groisman_microfluidic_2004,nguyen_improvement_2008,sousa_efficient_2010}. 
Among the few mechanisms employed to rectify Newtonian fluid flows under the Stokes regime, 
one can refer to mechanically intervening in the microchannel design 
using the flap structure 
\cite{adams_polydimethylsiloxane_2005,loverich_concepts_2006,loverich_single-step_2007}, 
or ball/particle-based check-valves 
\cite{pan_magnetically_2005,ou_microspheres_2012,sochol_microfluidic_2014}, 
where the flap and balls allow the fluid to 
preferentially flow in one direction.
In addition to the channel clogging risk due to the flap structure and ball/particle-based check-valves,  
a notable difficulty associated with these mechanisms is that 
the particles and biospecies to be transferred 
can get damaged or stuck after hitting the flap and balls/particles.
More importantly, to our knowledge, rectification of gaseous flows with equilibrium processes (Knudsen number smaller than 0.001) has not been demonstrated in the Stokes flow regime, warranting an alternative approach.



\begin{figure}[!b]
\centering
	\subfloat{%
	\includegraphics[width=0.35\textwidth]{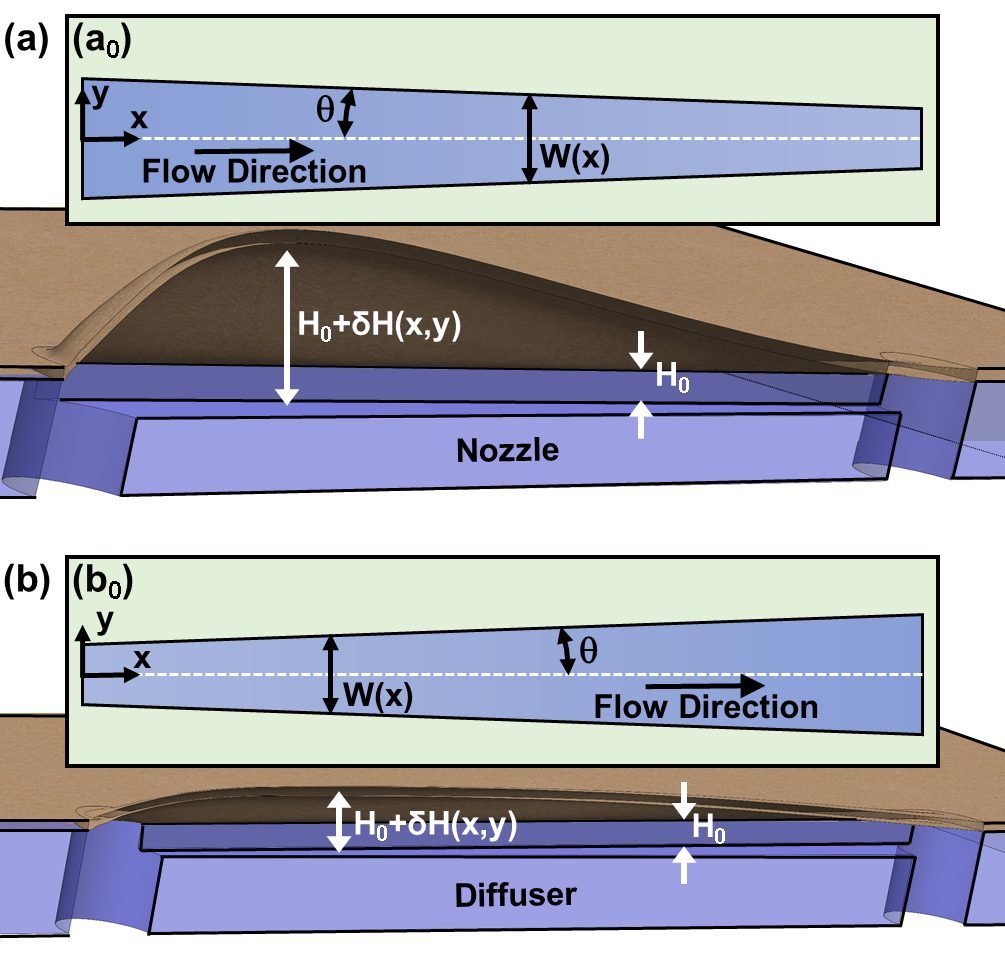}
	}
	\caption{Schematic diagrams showing the cross-section views of the fluid flow through a microchannel with a deformable ceiling in (a) nozzle and (b) diffuser directions together
 with (a\textsubscript{0} and b\textsubscript{0}) the top view of the channel.
The dimensions in z-direction
are not in scale with those in x-y plane for the sake of clarity.
	}
	\label{Fig-Schem}
\end{figure}

Here we present 
a more straightforward, yet universal, route
to introducing nonlinear effects to 
the equations of motion for Stokes flow of Newtonian fluids 
(for both liquids and gases)
by using the framework of deformable microchannels. 
We demonstrate that a shallow nozzle/diffuser microchannel with deformable ceiling provides a non-linear and direction-dependent governing coupled fluid-solid-mechanics equation leading to asymmetric hydrodynamic resistances enabling flow rectification of Newtonian fluids under the Stokes flow regime. 
The schematic representation of the fluid flow through a nozzle/diffuser microchannel with a deformable ceiling is shown in Fig.~\ref{Fig-Schem}.
The ceiling membrane deflection depends on the fluid flow direction, {\it i.e.} nozzle versus diffuser, causing the overall hydrodynamic resistances of diffuser and nozzle to differ from each other.
The one-dimensional model used in this work is based on the lubrication theory, {\it i.e.} 
${H\textsubscript{0}}\ll {W}\ll {L}$, 
where 
${H\textsubscript{0}}$, ${W}$, and ${L}$ 
refer to the microchannel's original height, 
width, and length, respectively.
The {thin-plate-bending framework} is considered for the elastic deformation, 
which requires 
$\delta H\ll {t}\ll {W}$, 
where $\delta H$ is the membrane displacement and ${t}$ denotes the membrane thickness.
Under these situations, 
the displacement of an arbitrary membrane's infinitesimal slice across the channel can be correlated solely with the local fluid pressure within the channel \cite{christov_flow_2018}.
The deflection profile of the slice is then obtained from the Euler-Bernoulli beam theory:
\begin{equation}
\frac{\delta H(x,\zeta)}{H_0}=\gamma(x)f(\zeta),
\label{Eq-wide-beam-deflection}
\end{equation}
where 
$\zeta\equiv 2y/W(x)$, $f(\zeta)=(\zeta+1)^2(\zeta-1)^2$,
$\gamma(x)= {p(x)W^4(x)}/{384DH_0}$,
and
$D = \frac{Et^3}{12(1-\nu^2)}$,
in which $E$ and $\nu$ denote the membrane's modulus of elasticity and Poisson's ratio 
\cite{roark_roarks_2002}.
In addition, $p$ refers to the gauge pressure within the channel, which is constant at any x-plane (normal to the x-axis) according to the lubrication theory.

Since the membrane displacement is significantly smaller than the channel width and length, {\it i.e.} $\delta H\ll {W}\ll {L}$,
the membrane slope is very small, {\it i.e.} $\partial \delta H/\partial x \ll 1$ and $\partial \delta H/\partial y \ll 1$.
Under these situations, and considering the sufficiently small nozzle/diffuser half-angle ($\theta$) with an attached flow throughout the channel, 
the local mass flow rate can be written as $\dot{m}(x)=-\frac{{dp(x)}/{dx}}{12\mu}{\rho(x)W(x)<H^3>_{(x)}}$,
where $\rho$ and $\mu$ refer to the fluid density and dynamic viscosity, respectively, and $H=H_0+\delta H$ shows the channel's local height.
In addition, the average of an arbitrary variable such as $\phi$ over a x-plane is 
obtained through $<\phi>_{(x)}=\frac{1}{2}\int^{\zeta=+1}_{\zeta=-1}{\phi(x,\zeta) d\zeta}$.
In a non-dimensional form, where the fluid density
at ambient pressure and ambient temperature, $\rho_\text{ref}$, is used for the density unit,
$L$ for the length unit, 
$\dot{m}_\text{ref}$ for the mass flow rate unit,  
$Q_\text{ref}=\dot{m}_\text{ref}/\rho_\text{ref}$ for the volumetric flow rate unit,  
and 
$\Delta p_\text{ref}\equiv {12\mu LQ_\text{ref}}/{W_iH^3_0}$
as the pressure unit, 
the dimensionless mass flow rate through the deformable channel, {\it i.e.} $\dot{m}^{\ast}\equiv {\dot{m}}/{\dot{m}_\text{ref}}$, can be written as
\begin{equation}
\dot{m}^{\ast}=-\rho^\ast\frac{dp^\ast(x^\ast)}{dx^\ast}\frac{W^\ast(x^\ast)}{W^\ast_i}\frac{<H^3>_{(x^\ast)}}{H^3_0},
\label{Eq-ND-Q}
\end{equation}
where $W_i$ refers to the channel width at inlet,
$\rho^\ast\equiv\rho/\rho_\text{ref}$,
$p^\ast\equiv p/\Delta p_\text{ref}$, $x^\ast\equiv x/L$, 
$W^\ast\equiv W/L$, and $W^\ast_i\equiv W_i/L$.
In general, the fluid density can be variable within the channel due to the compressibility effects.
For an isothermal fluid flow, we have
$\rho^\ast=1+p{\kappa_T}$, where $\kappa_T$ denotes the isothermal compressibility of the fluid at ambient pressure and ambient temperature.
Considering that 
$<H^3> = H^3_0+3H^2_0<\delta H>+3H_0<(\delta H)^2>+<(\delta H)^3>$,
one can show that
\begin{equation}
\begin{aligned}
\frac{<H^3>_{(x^\ast)}}{H^3_0}&=1+3\gamma(x^\ast)<f(\zeta)>_{(x^\ast)}\\
+3\gamma^2(x^\ast)&<f^2(\zeta)>_{(x^\ast)}+\gamma^3(x^\ast)<f^3(\zeta)>_{(x^\ast)}.
\label{Eq-H3_avg_per_H0_3}
\end{aligned}
\end{equation}

\noindent We know $<f(\zeta)>={8}/{15}$, $<f^2(\zeta)>={128}/{315}$, and $<f^3(\zeta)>={1024}/{3003}$. 
Considering that $p^\ast=0$ at $x^\ast=1$ due to the fluid discharge into the ambient air, 
after defining $\xi\equiv 1-x^\ast$ and substituting Eq.~\ref{Eq-H3_avg_per_H0_3} in Eq.~\ref{Eq-ND-Q},  
one can obtain an initial-value problem for an arbitrary mass flow rate of $\dot{m}=\dot{m}_\text{ref}$, {\it i.e.} $\dot{m}^{\ast}=1$, as shown in the following first-order non-linear ordinary differential equation (ODE):
\begin{equation}
\begin{aligned}
\Bigg(1+&\alpha_1\tau^4(\xi)p^\ast(\xi)+
\alpha_2\tau^8(\xi){p^\ast}^2(\xi)
+\alpha_3\tau^{12}{p^\ast}^3(\xi)\Bigg)\times\\
&\text{~~~~~~~}\Big(1+p^\ast(\xi){\kappa^\ast_T}\Big)\tau(\xi)\frac{dp^\ast(\xi)}{d\xi}=1,\\
&\text{~~~~~~Initial}\text{~value:\ \ \ \ \ }\xi=0~:~p^\ast=0,
\end{aligned}
\label{Eq-Final-ODE}
\end{equation}
where 
the dimensionless isothermal compressibility of the fluid is defined as 
$\kappa^\ast_T\equiv {{\kappa_T}}{\Delta p_\text{ref}}$,
and
the width profile function, {\it i.e.} $\tau(\xi)$, {flexibility parameter}, {\it i.e.} $\chi$, and $\alpha_i$ coefficients are
\begin{subequations}
\begin{align}
\tau(\xi)&\equiv {W(\xi)}/{W_i}\text{,}\\
\chi&\equiv \frac{\Delta p_\text{ref}W^4_i}{384DH_0}\text{,}\\
\alpha_1=\frac{8}{5}\chi \text{,~} \alpha_2&=\frac{128}{105}\chi^2 \text{, and~} \alpha_3=\frac{1024}{3003}\chi^3\text{.}
\end{align}
\label{Eq-Definitions_00}
\end{subequations}
For a nozzle/diffuser microchannel 
with a linear variation of the width, 
$\tau(\xi)=\tau_o+(1-\tau_o)\xi$, 
where $\tau_o\equiv W_o/W_i$ and $W_o$ refers to the channel width at outlet.
The governing equation is reduced to that of the fluid flow through a deformable shallow straight microchannel \cite{mehboudi_one-dimensional_2018} for the special case of $\tau=1$.
The ODE presented in Eq.~\ref{Eq-Final-ODE} is solved numerically to find the pressure distribution within the channel, {\it i.e.} $p^\ast(\xi)$, and other fluid-solid characteristics.

\begin{figure*}[!t]
\centering
	\subfloat{%
	\includegraphics[width=0.8\textwidth]{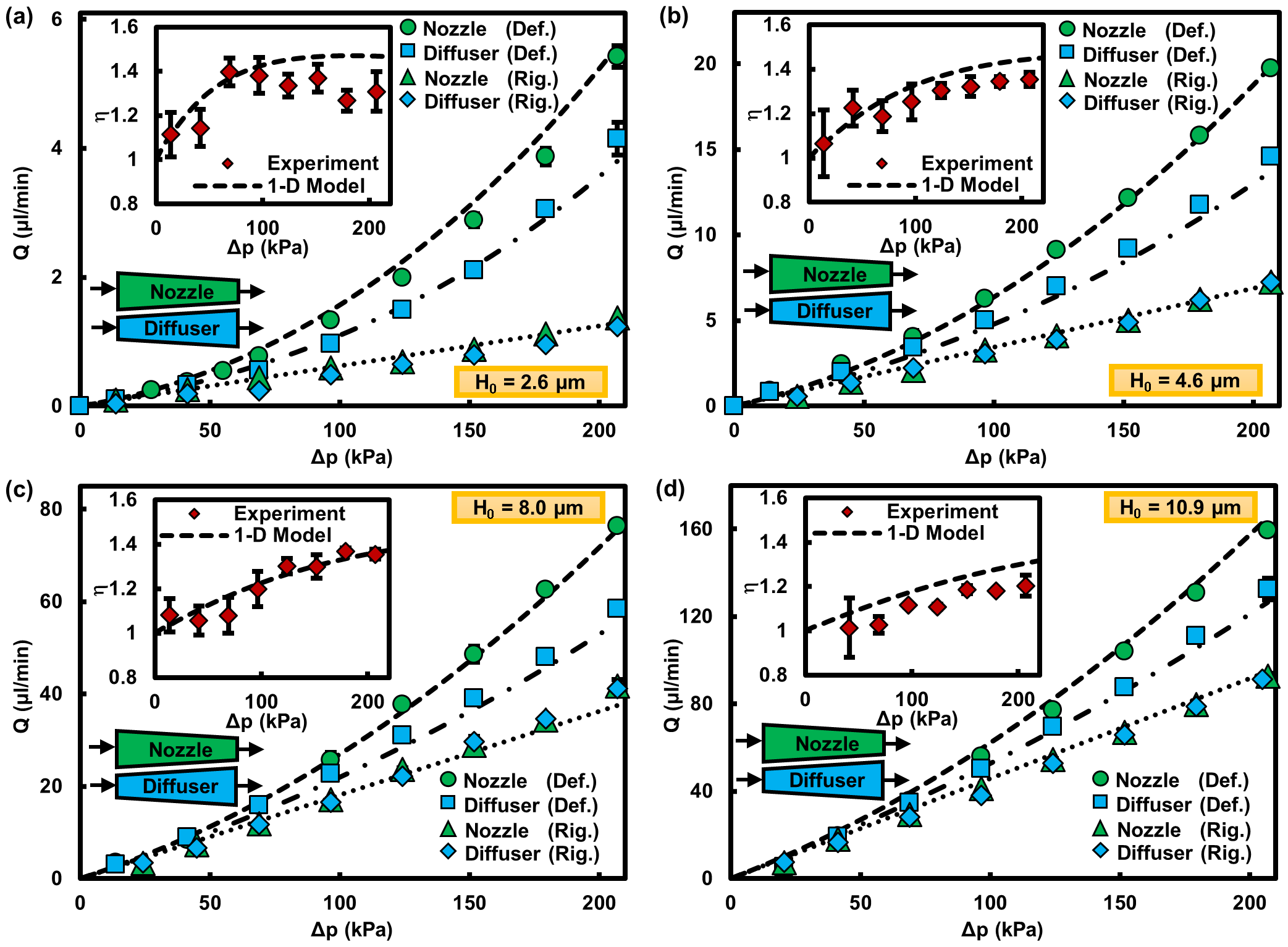}
	}
	\caption{Volumetric flow rates of DI water as a function of pressure difference across the deformable (Def.) and rigid (Rig.) nozzle/diffuser microchannels for various original heights of $H_0=$ (a) $2.6\pm 0.2~\mu m$, (b) $4.6\pm 0.2~\mu m$, (c) $8.0\pm 0.3~\mu m$, and (d) $10.9\pm 0.3~\mu m$.
	The dashed, dash-dotted, and dotted lines show the results obtained from the one-dimensional model for deformable nozzle, deformable diffuser, and rigid nozzle/diffuser, respectively.
	The insets show the rectification ratio 
	as a function of pressure difference across the deformable nozzle/diffuser microchannels.
	The error bars related to the experimental results (averaged over three runs) are not visible if they are smaller than the markers' size.
	}
	\label{Fig-Res-Height-Effects}
\end{figure*}
To validate the mathematically-described underlying coupled fluid-solid mechanics of fluid flow through an asymmetrically-shaped deformable microchannel, we fabricated microfluidic devices and performed experiments to obtain the flow rate over the applied pressure difference range of 14--206 kPa.
The large and small widths of the fabricated nozzle/diffuser channels equal $2~mm$ and $1~mm$, respectively. 
The half-angle is $\theta=1.25^\circ$, {\it i.e.} $L\approx 22.9~mm$. 
The half-angle is sufficiently small so that $dW/2dx\approx 0.022$ meets the requirement of $dW/2dx\ll 1$ in our model. 
The obtained experimental results are shown in Fig.~\ref{Fig-Res-Height-Effects} for DI water 
flow through deformable/rigid nozzle/diffuser microchannels with different original heights.
The calculated Reynolds number in the mid-plane ($x^\ast=0.5$) varies from $O(10^{-3})$ to $O(10^{0})$. 
Because of the linear nature of equations governing the low-Reynolds-number flows through rigid channels, rigid nozzle and diffuser exhibit the same behavior resulting in no flow rectification. 
For the deformable microchannels, however, because of the membrane deformation 
the deformable nozzle delivers a larger mass flow rate compared to the deformable diffuser under the same pressure difference resulting in a flow rectification.
The rectification ratio, {\it i.e.}  
$\eta\equiv \dot{m}_\text{Nozzle}/\dot{m}_\text{Diffuser}$, 
approaches unity as the pressure difference decreases toward zero. 
A small pressure difference means a small flexibility parameter, thus diminishing
the role of non-linear terms in the governing coupled fluid-solid-mechanics equation shown in Eq.~\ref{Eq-Final-ODE}. 
As a result, a deformable microchannel exhibits a characteristic behavior similar to that of its rigid counterpart without flow rectification.
The analytical solution presented in Supplemental Material A.1 confirms this observation.
A sufficiently large increase in pressure difference activates the non-linear terms of $\alpha_i\tau^{4i}(\xi){p^\ast}^i(\xi)~:~i=1,~2,$ and $3$ in 
Eq.~\ref{Eq-Final-ODE}
allowing the embedded non-linear and direction-dependent nature of the governing coupled fluid-solid-mechanics equation to emerge, because of which the flow rectification is observed.

\begin{figure}[!t]
\centering
	\subfloat{%
	\includegraphics[width=0.4\textwidth]{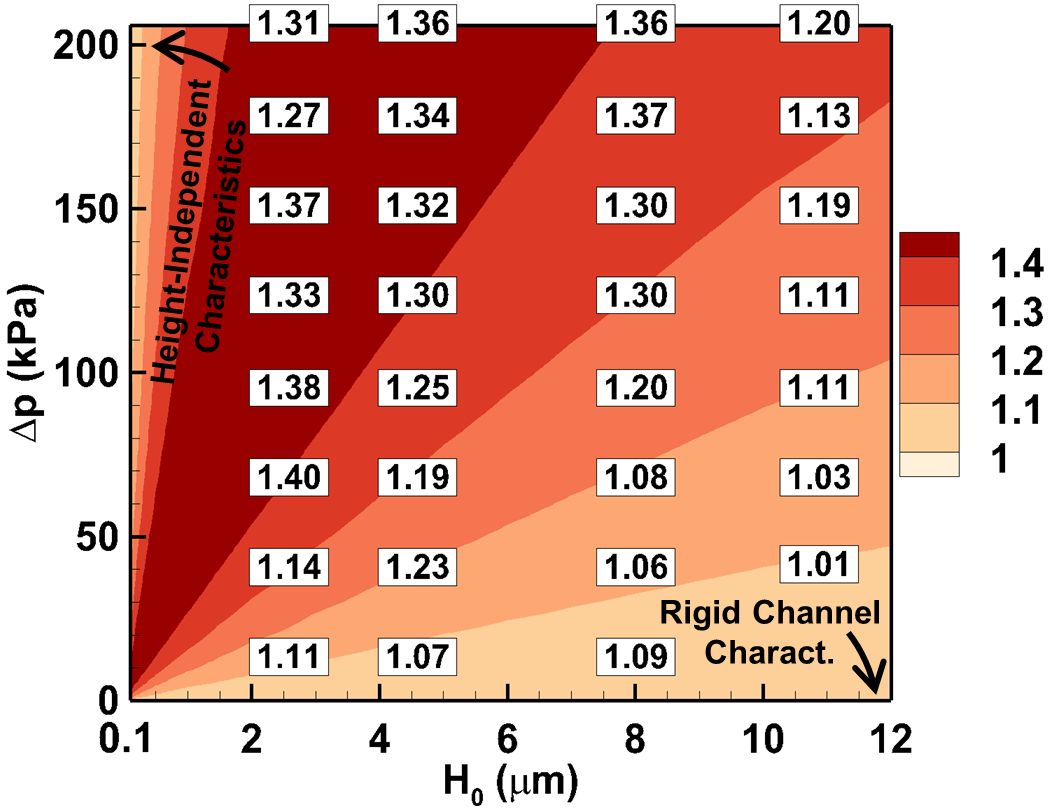}
	}
	\caption{The rectification ratio contour obtained from the 1-D model for DI water flow through the nozzle/diffuser microchannels with the 
	widths of $1~mm$ and $2~mm$ together with the experimental results 
	tabulated at the corresponding  
	($H_0,\Delta p$).
	}
	\label{Fig-Res-Contour}
\end{figure}

From Fig.~\ref{Fig-Res-Height-Effects}, increasing the channel's original height can attenuate the contrast between the deformable and rigid channels' characteristics, 
since 
the membrane deformation becomes less significant compared with the channel's original height,
and consequently
the membrane deformation-induced change in hydrodynamic resistance plays a less noticeable role in the net hydrodynamic resistance. 
Consequently, a rectification ratio decreases as the channel's original height increases.
Meanwhile, 
a rectification ratio decreases after reaching its maximum with increasing the pressure difference; see Fig.~\ref{Fig-Res-Contour}.
The analytical solution presented in Supplemental Material A.2 
for the case of sufficiently large pressure difference, where the third order of flexibility parameter ($\chi^3$) dictates the behavioral characteristics, 
confirms this observation, as well as demonstrating a height-independent characteristic behavior.

Our experimental and theoretical investigations also prove the validity of the derived model and emergence of the flow rectification for other incompressible and compressible flows by using Methanol, Isopropyl alcohol, and air as the working fluids; see Supplemental Material B.
Even though the flow rate decreases as the fluid viscosity increases, the rectification ratio remains the same for different incompressible flows. 
The reason is that for a given microchannel, the dimensionless flow rate and consequently the rectification ratio depend on the dimensionless pressure and dimensionless density distributions within the nozzle/diffuser microchannels (Eq.~\ref{Eq-ND-Q}), which are solely dictated by the fluid compressibility, membrane properties as well as the channel's width and original height; see Eqs.~\ref{Eq-Final-ODE} and \ref{Eq-Definitions_00}. 

The model derived in this work simultaneously takes channel deformability and fluid compressibility into account, 
leading to the universal rectifier for Stokes flows of both liquids and gases.
We found that 
the rectification ratio of air flow is slightly smaller than that of the liquid flows (Supplemental Material B). 
For the flow of a fluid with larger compressibility, the local fluidic resistance increases along the channel because of a larger density gradient within the channel.
The fluid density decreases along the channel as the pressure decreases along the channel.
According to the conservation of mass principle, such a decaying density variation causes the fluid speed, shear rate, local frictional resistance, and net fluidic resistance to increase along the channel.
As a result, the contribution of the wide portion in the nozzle's  hydrodynamic resistance decreases for a nozzle, while that increases for a diffuser. 
Consequently, the membrane displacement-induced reduction in the hydrodynamic resistance of a nozzle decreases, while that increases for a  diffuser, leading to a rectification ratio smaller than that of the incompressible flows. 
Regardless of the lower rectification ratios of compressible flows compared with those of incompressible flows, this work still reports the first demonstration of rectifying an equilibrium gas flow (Knudsen number smaller than 0.001) under the Stokes flow regime with rectification ratios of $\sim$ 1.4.

\begin{figure*}[!t]
\centering
	\subfloat{%
	\includegraphics[width=0.8\textwidth]{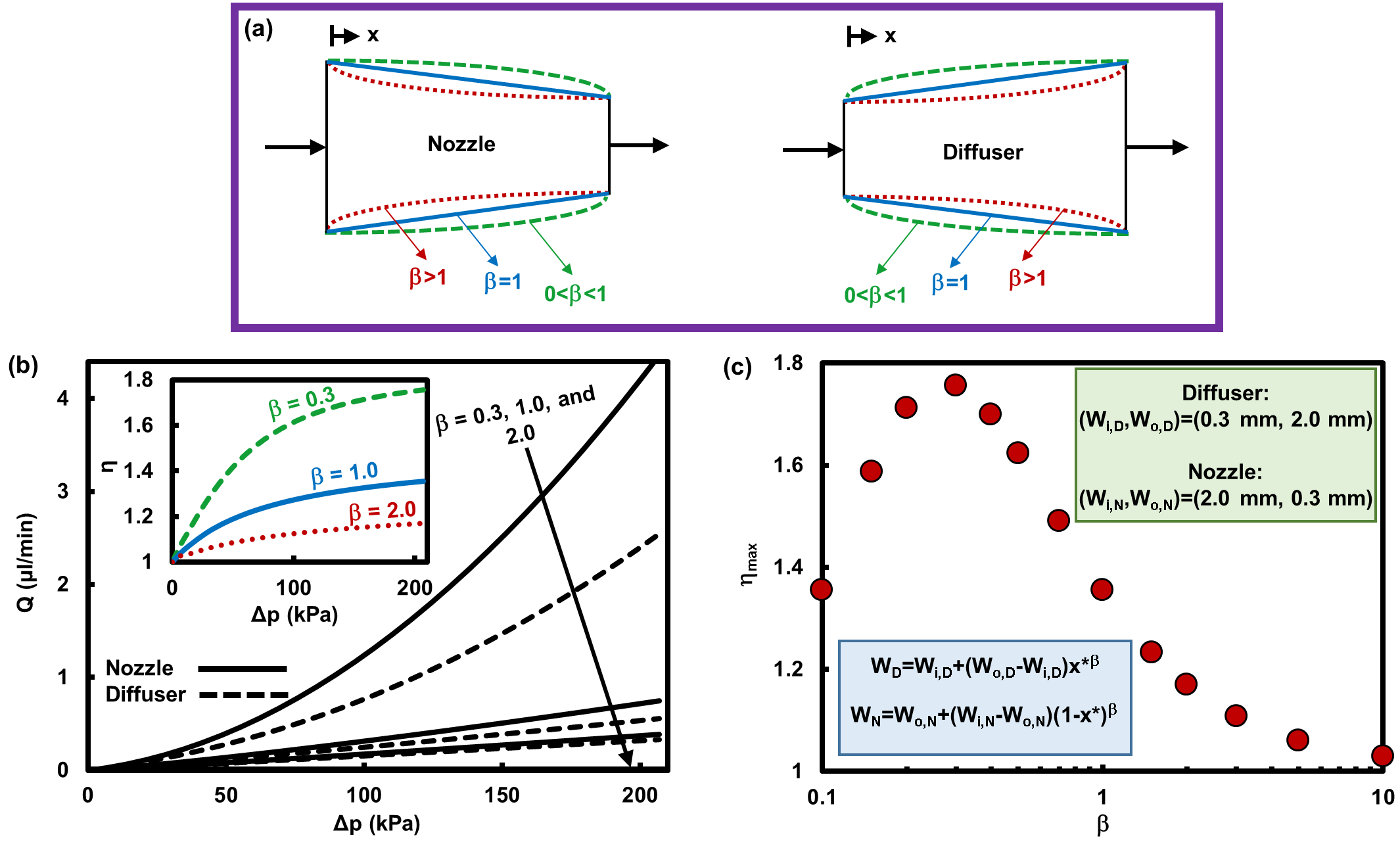}
	}
	\caption{(a) Schematic presentation of the nonlinear microchannel width profiles in the nozzle direction, {\it i.e.} $W_N=W_o+(W_i-W_o)(1-x^\ast)^\beta$, as well as the diffuser direction, {\it i.e.} $W_D=W_i+(W_o-W_i){x^\ast}^\beta$.
	(b) Volumetric flow rates of water obtained from the 1-D coupled fluid-solid-mechanics model as a function of pressure difference across the deformable nozzle/diffuser microchannels of $2.6~\mu m$ in height with the half-angle of $\theta=1.25^\circ$, and small and large widths of $0.3~mm$ and $2~mm$ for different width profiles.
	The inset shows the rectification ratio, {\it i.e.} $\eta=Q_\text{Nozzle}/Q_\text{Diffuser}$, as a function of pressure difference across the channel.
	(c) The maximum rectification ratio for microchannels with different width profiles over the pressure difference range of 
	14 kPa -- 206 kPa. 
	}
	\label{Fig-Width_Profile}
\end{figure*}


Passive rectifiers with improved performances can be realized through optimizing the width profile.
Extending the wide portion of the asymmetric microchannel using appropriate nonlinear width profiles intensifies the contribution of the wide portion in the microchannel's  hydrodynamic resistance. 
As a result, the membrane displacement-induced change in the hydrodynamic resistance becomes more comparable with the overall hydrodynamic resistance, leading to larger rectification ratios. 
For example, the maximum rectification ratio of the deformable nozzle/diffuser microchannel of $2.6~\mu m$ in height presented in Fig.~\ref{Fig-Res-Height-Effects} can be improved from $\sim 1.43$ for a linear width profile ($\beta=1$) to $\sim 1.76$ for a modified nonlinear one with $\beta=$ 0.3; see Fig.~\ref{Fig-Width_Profile}. 


We have shown that 
passive rectification of compressible and incompressible flows under the Stokes regime is universally possible for Newtonian fluids (liquids and gases) by introducing
the non-linear and direction-dependent terms to the 
otherwise linear equations of motion. 
The proposed nonlinearity stems from the coupled fluid-solid mechanics 
in an asymmetrically-shaped microchannel 
with deformable ceiling.
An important advantage of the proposed rectification mechanism 
is that the microchannel can be reliably designed using the derived analytical model.
The demonstrated rectification ratio of $\sim$ 1.2--1.8 is comparable with many of the rectifiers resorting to non-Newtonian fluids under the Stokes flow regime \cite{groisman_microfluidic_2003,groisman_microfluidic_2004,nguyen_improvement_2008}.
The proposed platform is also more compatible with biospecies transport as opposed to the rectifiers 
using the flap structure \cite{adams_polydimethylsiloxane_2005,loverich_concepts_2006,loverich_single-step_2007}, 
or ball/particle-based check-valves \cite{ou_microspheres_2012,sochol_microfluidic_2014}, wherein the biospecies can be damaged by or stuck to the flap or particles/balls.
Furthermore, due to the large hydrodynamic resistance of such shallow rectifiers, the leakage flow rate under the off mode is significantly small, 
suggesting the described underlying physics may be exploited to 
develop precise micropumps with low flow-rates as well as small-leakage microvalves with potential bioapplications such as species transport, drug delivery, etc.
Finally, we note that rectifiers with smaller dimensions can be made by using shorter and narrower microchannels sealed by a more compliant membrane, which suggests new opportunities for microfluidic integrated circuits.


\bibliographystyle{apsrev4-1}

\bibliography{DND_01}


\section*{Acknowledgments}
\label{Acknowledgments} 

Authors thank Dr. Baokang Bi and staff of W. M. Keck Microfabrication Facility, and Karl Dersch and staff of ECE Research Cleanroom in Michigan State University for their assistance in the cleanroom.


\section*[Methods]{Methods}
\label{Methods}
\noindent {\bf Fabrication.}
The microchannels were created through glass wet etching process, 
drilling the via-holes,
and adhesive bonding using a $2~\mu m$-SU8-coated PET (polyethylene terephthalate) film 
($t=100\pm 5~\mu m$, $E=2\text{--}4$ GPa, and $\nu\approx 0.4$) 
followed by bonding reinforcement using a glass slide with a sufficiently deep ($\sim 45~\mu m$) etched pattern of the microchannel allowing the membrane to deform freely under pressure. 
The reinforcement glass was bonded to the bare side of the membrane using a $16~\mu m$ SU8 layer ($E\sim 2$ GPa) 
\cite{mehboudi_experimental_nodate}.
The blank glass slides were used for reinforcement of rigid channels to avoid membrane deformations \cite{mehboudi_two-step_2018}.

\noindent {\bf Capturing adhesive layers in the model.}
The structural effects of the adhesive layers are taken into account in our model by using the transformed-section method \cite{gere_mechanics_2004}. 
The equivalent membrane thickness is calculated from 
$\tilde{t}=\sqrt[3]{12I/W}$, where $I$ denotes the three-layer membrane's moment of inertia. 
The equivalent membrane thickness, {\it i.e.} $\tilde{t}$, replaces $t$ in the parameter $D$ of Eq.~\ref{Eq-Definitions_00} (b) to evaluate the flexibility parameter.
The modulus of elasticity of PET film was considered 3 GPa and 3.5 GPa for modeling the liquid and gas flows, respectively. 

\noindent {\bf Experiments.}
{\it Liquid as the working fluid:} A pressure-driven flow was generated at constant pressure levels by employing a compressed air to push a working fluid out of a custom-made flask with two ports. A pressure source is connected to one end of the flask via a pressure regulator (PneumaticPlus, PPR2-N02BG-4 Miniature Air Pressure Regulator), while the other end is connected to the microchannel's inlet.   
The liquid discharged from the microchannel was guided through a tube with a known inner diameter, where the meniscus position was monitored over time to calculate the volumetric flow rate.\\
{\it Gas as the working fluid:} Air was chosen as a model compressible fluid. The in-line pressure of air flow was adjusted using a pressure regulator (PneumaticPlus, PPR2-N02BG-4 Miniature Air Pressure Regulator). A small water plug was placed within the outlet tube and pushed out by the applied air pressure. The movement of water plug was monitored with a camera attached to a stereomicroscope, and linear velocity of liquid meniscus was measured to calculate the volumetric flow rate of air at outlet and consequently the mass flow rate through the channel. 

\noindent {\bf Properties of the fluids.}
The ambient temperature and pressure are $T_\text{atm}=295$ K and $p_\text{atm}=101$ kPa, respectively.
The dynamic viscosity of air is considered to be the constant value of $\mu=1.8\times 10^{-5}$ Pa.s.
The air specific gas constant equals $R_\text{sgc}=287.058$ J.kg\textsuperscript{-1}.K\textsuperscript{-1}.
Air is considered as an ideal gas in this work, the isothermal compressibility of which is $\kappa_T={1}/{p_\text{atm}}$.
The dynamic viscosity of DI water, Methanol and Isopropyl Alcohol is $8.9\times 10^{-4}$ Pa.s, $5.9\times 10^{-4}$ Pa.s, and $2.3\times 10^{-3}$ Pa.s, respectively. 
The isothermal compressibility of liquids is considered to be $\kappa_T=0$, {\it i.e.} incompressible fluid flows.
Also, according to Eq.~\ref{Eq-Definitions_00}, the flexibility parameter and consequently the $\alpha_i$ coefficients in Eq.~\ref{Eq-Final-ODE} are zero for fluid flow through a rigid microchannel.


\clearpage
\newpage

\pagestyle{Appendix_A}

\makeatletter\onecolumngrid@push\makeatother

{\centerline{\bf{\large{A passive Stokes flow rectifier for Newtonian fluids}}}}
{\centerline{\large Aryan Mehboudi and Junghoon Yeom}}
{\centerline{Department of Mechanical Engineering, }}
{\centerline{428 South Shaw Ln., East Lansing, MI 48824, US.}}
{\centerline{{\it Email address:} jyeom@egr.msu.edu}}

\section*{A. Studying special cases analytically}
\subsection*{A.1. Extremely small flexibility parameter}
\label{Rigid}

\begin{figure}[!bt]
\centering
	\subfloat{%
	\includegraphics[width=0.4\textwidth]{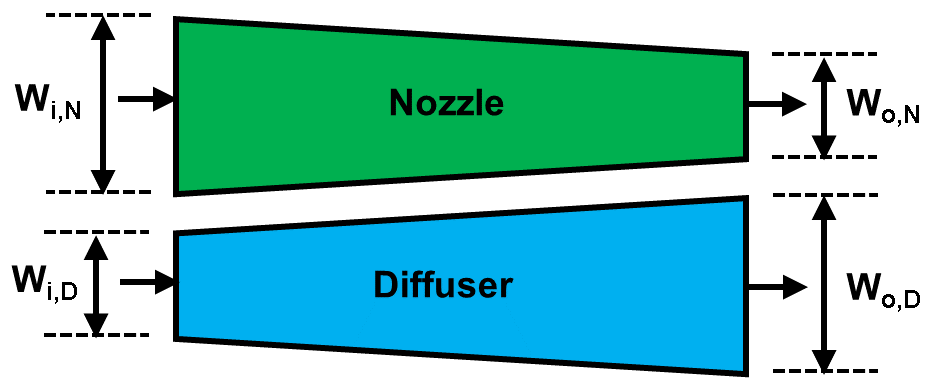}
	}
	\caption{Diagram showing the nomenclature of the nozzle/diffuser inlet/outlet dimensions.
	}
	\label{Fig-ND-Schem-Dimensions}
\end{figure}

For a rigid channel, the flexibility parameter equals zero, {\it i.e.} $\chi=0$.
For deformable channels under sufficiently small pressure differences, the flexibility parameter can also be very small, {\it i.e.} $\chi\approx0$.
Under these conditions, the terms with $\alpha_i~:~i=1,2,$ and $3$ multipliers vanish from the coupled fluid-solid-mechanics governing equation presented in Eq.~\ref{Eq-Final-ODE}, simplifying the ODE into
\begin{equation}
\begin{aligned}
\tau(\xi)\Big(1+p^\ast(\xi){\kappa^\ast_T}\Big)\frac{dp^\ast(\xi)}{d\xi}&=1,\\
\text{Initial}\text{~value:\ \ \ \ \ }\xi=0~:~p^\ast&=0.
\end{aligned}
\label{Eq-Final-ODE-Small-Flexibility}
\end{equation}
For a nozzle/diffuser, where $\tau(\xi)=\tau_o+(1-\tau_o)\xi$, one can solve Eq.~\ref{Eq-Final-ODE-Small-Flexibility} to obtain the following analytical expression for the pressure distribution within the channel:
\begin{equation}
\begin{aligned}
p^\ast(\xi)=\frac{1}{\kappa^\ast_T}\Bigg(\sqrt{1+\frac{2\kappa^\ast_T}{1-\tau_o}\ln\frac{\tau(\xi)}{\tau_o}}-1\Bigg).
\end{aligned}
\label{Eq-Press-Dist-Rig-00}
\end{equation}
The dimensionless pressure at inlet can then be obtained as
\begin{equation}
\begin{aligned}
p^\ast_i=\frac{1}{\kappa^\ast_T}\Bigg(\sqrt{1+{2\kappa^\ast_T}\frac{\ln{\tau_o}}{\tau_o-1}}-1\Bigg).
\end{aligned}
\label{Eq-Press-Inlet-Rig-00}
\end{equation}
Since 
$\Delta p=p^\ast_i \Delta p_\text{ref}$,
$\Delta p_\text{ref}\equiv {12\mu L\dot{m}_\text{ref}}/{W_iH^3_0\rho_\text{ref}}$, 
and 
$\dot{m}=\dot{m}_\text{ref}$, {\it i.e.} $\dot{m}^{\ast}=1$,
the mass flow rate correlation with pressure difference across the channel can be elicited as  
\begin{equation}
\begin{aligned}
\dot{m}=\bigg(\frac{\tau_o-1}{2\kappa_T\ln(\tau_o)}\bigg)\bigg(\frac{W_iH^3_0\rho_\text{ref}}{12\mu L}\bigg)\bigg(\big(1+\kappa_T\Delta p\big)^2-1\bigg).
\end{aligned}
\label{Eq-Q-P-Rig-00}
\end{equation}
As a reminder, $\rho_\text{ref}$ is the fluid density at ambient pressure and ambient temperature.
Using Eq.~\ref{Eq-Q-P-Rig-00}, we can calculate the rectification ratio as follows:
\begin{equation}
\begin{aligned}
\eta\equiv\frac{\dot{m}_\text{Nozzle}}{\dot{m}_\text{Diffuser}}=\frac{\frac{\tau_{o,N}-1}{\ln(\tau_{o,N})}\times W_{i,N}}{\frac{\tau_{o,D}-1}{\ln(\tau_{o,D})}\times W_{i,D}},
\end{aligned}
\label{Eq-Rectification-Rig-00}
\end{equation}
where $W_{i,N}$ and $W_{i,D}$ refer to the width at inlet section for the fluid flow in nozzle and diffuser directions, respectively.
Similarly,   
$W_{o,N}$ and $W_{o,D}$ denote the width at outlet section of for the fluid flow in nozzle and diffuser directions, respectively.
We have $\tau_{o,N}=W_{o,N}/W_{i,N}$ and $\tau_{o,D}=W_{o,D}/W_{i,D}$.
We know that $W_{i,N}=W_{o,D}$, $W_{i,D}=W_{o,N}$, and
$\tau_{o,N}=1/\tau_{o,D}$; see Fig.~\ref{Fig-ND-Schem-Dimensions}.
Since $\ln(\tau_{o,N})=-\ln(\tau_{o,D})$, after some algebraic calculations, one can obtain
\begin{equation}
\begin{aligned}
\eta&=\frac{\tau_{o,N}-1}{1-\tau_{o,D}}\frac{W_{i,N}}{W_{i,D}}\\
&=\frac{\tau_{o,N}-1}{1-1/\tau_{o,N}}\frac{1}{\tau_{o,N}}\\
&=1,
\end{aligned}
\label{Eq-Rectification-Rig-01}
\end{equation}
which states there is no flow rectification for Newtonian fluids (liquid/gas) using a rigid nozzle/diffuser microchannel under the Stokes flow regime.

The relation for a straight channel, {\it i.e.} $\tau_o=1$, can be obtained through $\lim_{\tau_o \to 1}\dot{m}$, that results in
\begin{equation}
\begin{aligned}
\dot{m}=\bigg(\frac{W_iH^3_0\rho_\text{ref}\Delta p}{12\mu L}\bigg)\bigg(1+\frac{\kappa_T}{2}\Delta p\bigg),
\end{aligned}
\label{Eq-mdot-P-Large-Flex-Straight-00-Comp}
\end{equation}
which is in agreement with our other work investigating the effects of channel deformability and fluid compressibility simultaneously \cite{mehboudi_one-dimensional_2018}.

\subsection*{A.2. Extremely large flexibility parameter}
\label{EL flexibility}

For deformable channels under sufficiently large pressure differences, the flexibility parameter can be extremely large, 
so that the term with the third order of flexibility parameter, {\it i.e.} $\chi^3$,
dominates the left-side of Eq.~\ref{Eq-Final-ODE} simplifying the ODE into
\begin{equation}
\begin{aligned}
\alpha_3\tau^{13}(\xi)\Big(1+p^\ast(\xi){\kappa^\ast_T}\Big){p^\ast}^3(\xi)\frac{dp^\ast(\xi)}{d\xi}&=1,\\
\text{Initial}\text{~value:\ \ \ \ \ }\xi=0~:~p^\ast&=0.
\end{aligned}
\label{Eq-Final-ODE-Large-Flexibility}
\end{equation}
For a nozzle/diffuser, where $\tau(\xi)=\tau_o+(1-\tau_o)\xi$, one can rearrange Eq.~\ref{Eq-Final-ODE-Large-Flexibility} as 
\begin{equation}
\begin{aligned}
\alpha_3\tau^{13}(1-\tau_o)\Big(1+p^\ast{\kappa^\ast_T}\Big){p^\ast}^3\frac{dp^\ast}{d\tau}&=1,\\
\text{Initial}\text{~value:\ \ \ \ \ }\tau=\tau_o~:~p^\ast&=0.
\end{aligned}
\label{Eq-Final-ODE-Large-Flexibility-1}
\end{equation}
The analytical solution can be then written as
\begin{equation}
\begin{aligned}
\bigg(\frac{\kappa^\ast_T}{5}{p^\ast}^5(\xi)+\frac{1}{4}{p^\ast}^4(\xi)\bigg)=\frac{\tau^{12}(\xi)-\tau_o^{12}}{12\tau_o^{12}\tau^{12}(\xi)\alpha_3(1-\tau_0)}.
\end{aligned}
\label{Eq-Press-Dist-Large-Flex-00-Comp}
\end{equation}
Because of the difficulties associated with finding the roots of the fifth-polynomial above, we consider the special case of incompressible flows ($\kappa_T\to 0$) to obtain the following analytical expression for the pressure distribution within the channel:
\begin{equation}
\begin{aligned}
p^\ast(\xi)=\bigg(\frac{\tau^{12}(\xi)-\tau_o^{12}}{3\tau_o^{12}\tau^{12}(\xi)\alpha_3(1-\tau_0)}\bigg)^\frac{1}{4}.
\end{aligned}
\label{Eq-Press-Dist-Large-Flex-00}
\end{equation}
The dimensionless pressure at inlet, where $\tau(\xi=1)=1$, can then be obtained as
\begin{equation}
\begin{aligned}
p^\ast_i=\bigg(\frac{1-\tau_o^{12}}{3\tau_o^{12}\alpha_3(1-\tau_0)}\bigg)^\frac{1}{4}.
\end{aligned}
\label{Eq-Press-Inlet-Large-Flex-00}
\end{equation}
Since 
$\alpha_3=\frac{1024}{3003}({\Delta p_\text{ref}W^4_i}/{384DH_0})^3$, 
$\Delta p=p^\ast_i \Delta p_\text{ref}$, and
$\Delta p_\text{ref}\equiv {12\mu LQ_\text{ref}}/{W_iH^3_0}$, 
the volumetric flow rate correlation with pressure difference across the channel, in the case of incompressible flows, can be elicited as 
\begin{equation}
\begin{aligned}
Q=\bigg(\frac{1}{664,215,552}\frac{(1-\tau_0)\tau_0^{12}}{1-\tau_0^{12}}\frac{W_i^{13}}{\mu LD^3}\bigg)\times \Delta p^4,
\end{aligned}
\label{Eq-Q-P-Large-Flex-00}
\end{equation}
which shows a height-independent characteristic behavior, because the membrane deformation is significantly larger than the original height of microchannel under this regime.
The relation for a straight channel, {\it i.e.} $\tau_o=1$, can be obtained through $\lim_{\tau_o \to 1}Q$, that results in
\begin{equation}
\begin{aligned}
Q=\frac{1}{664,215,552}\times\frac{W^{13}}{12\mu LD^3}\times \Delta p^4,
\end{aligned}
\label{Eq-Q-P-Large-Flex-Straight-00}
\end{equation}
which is in agreement with our other work investigating the channel deformability effects on characteristics of incompressible flow through straight microchannels  \cite{mehboudi_experimental_nodate}.
Using Eq.~\ref{Eq-Q-P-Large-Flex-00}, the rectification ratio of incompressible flows through a nozzle/diffuser channel under this regime is calculated as follows:
\begin{equation}
\begin{aligned}
\eta=\frac{Q_\text{Nozzle}}{Q_\text{Diffuser}}&=\frac{\frac{(1-\tau_{0,N})\tau_{0,N}^{12}}{1-\tau_{0,N}^{12}}\times W^{13}_{i,N}}{\frac{(1-\tau_{0,D})\tau_{0,D}^{12}}{1-\tau_{0,D}^{12}}\times W^{13}_{i,D}}\\
&=\frac{\frac{(1-\tau_{0,N})\tau_{0,N}^{12}}{1-\tau_{0,N}^{12}}}{\frac{(1-\tau_{0,N}^{-1})\tau_{0,N}^{-12}}{1-\tau_{0,N}^{-12}}}\times \frac{1}{\tau_{0,N}^{13}}\\
&=1.
\end{aligned}
\label{Eq-Rectification-Large-Flex-00}
\end{equation}

\clearpage
\newpage
\pagestyle{Appendix_B}

\section*{B. Other working fluids}
\label{Supp_Res}


\subsection*{B.1. Liquid -- Incompressible flow}
\label{Supp_Res_Other_Liquid_flows}

\begin{figure}[!hbt]
\centering
	\subfloat{%
	\includegraphics[width=0.45\textwidth]{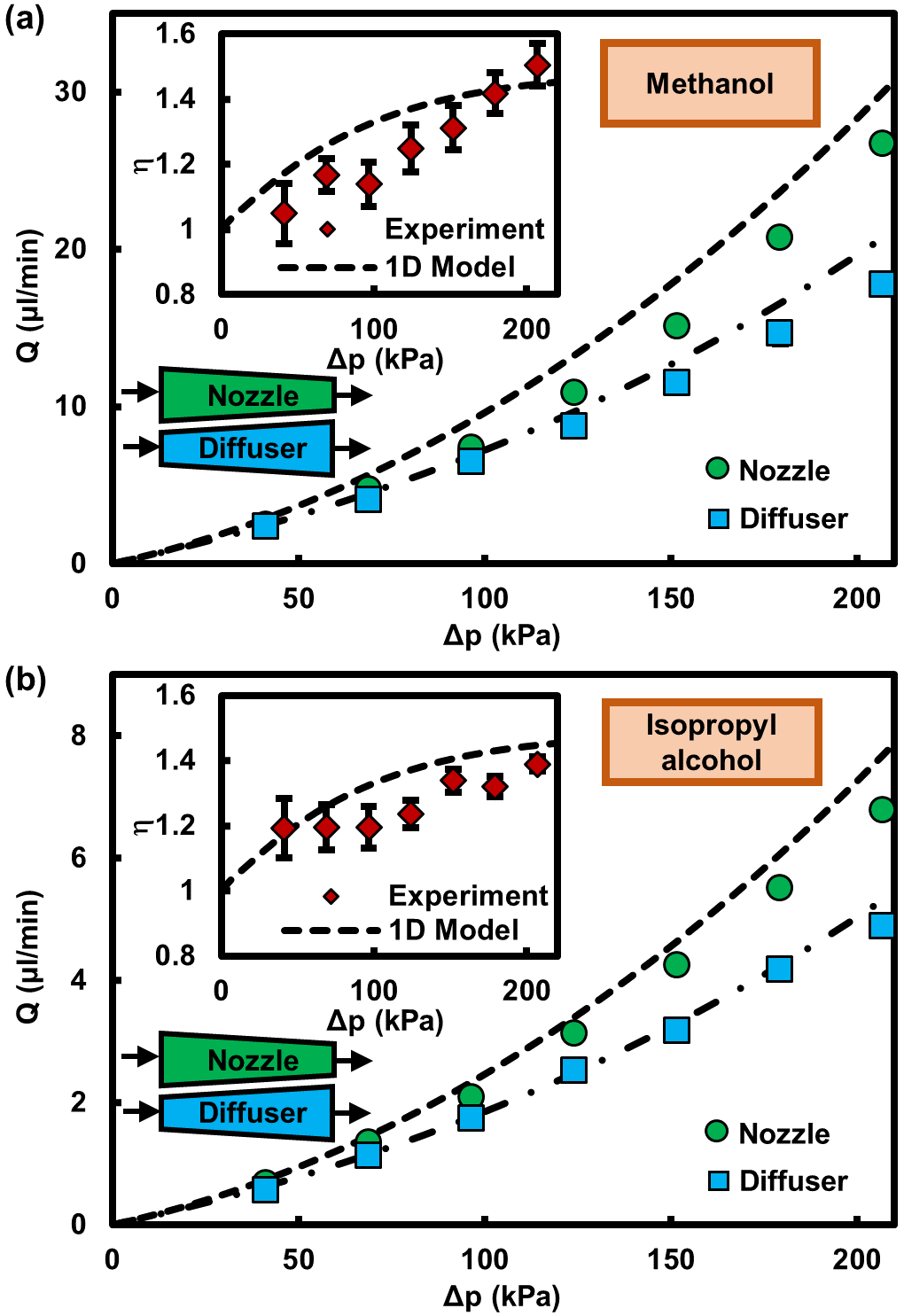}
	}
	\caption{Volumetric flow rate as a function of pressure difference across the deformable nozzle/diffuser microchannel of $\sim 4.6~\mu m$ in height for two different working fluids: (a) Methanol and (b) Isopropyl alcohol.
	The dashed and dash-dotted lines show the results obtained from the one-dimensional model for the deformable nozzle and diffuser, respectively.
	The insets show the rectification ratio 
	as a function of pressure difference across the channel.
	}
	\label{Fig-Res-Viscosity-Effects}
\end{figure}
To investigate the viscosity effects, we experimented with different working fluids whose viscosity is substantially different from that of DI water. Methanol ($\mu=5.9\times 10^{-4}$ Pa.s) and Isopropyl Alcohol ($\mu=2.3\times 10^{-3}$ Pa.s) have been tested in the deformable nozzle/diffuser with $4.6~\mu m$ in height. Figure~\ref{Fig-Res-Viscosity-Effects} depicts the theoretical and experimental results of the volumetric flow rate versus pressure drop as well as the flow rectification ratios. 
As explained in the text, the fluid viscosity influences the flow rate, but it does not affect the rectification ratio.
The dimensionless flow rate and consequently the rectification ratio depend on the dimensionless pressure and dimensionless density distributions within the nozzle/diffuser microchannels (Eq.~\ref{Eq-ND-Q}). 
Dimensionless density is unity within channel for all incompressible flows.
Dimensionless pressure distribution of incompressible flows is also solely dictated by the membrane properties and channel's geometry; see Eqs.~\ref{Eq-Final-ODE} and \ref{Eq-Definitions_00}. 
As a result, under the same pressure difference across the channel, a rectifier results in the same rectification ratio for all incompressible flows of Newtonian fluids under the Stokes flow regime.

\subsection*{B.2. Gas -- Compressible flow}
\label{Supp_Res_Gas_flows}

\begin{figure*}[!t]
\centering
	\subfloat{%
	\includegraphics[width=0.8\textwidth]{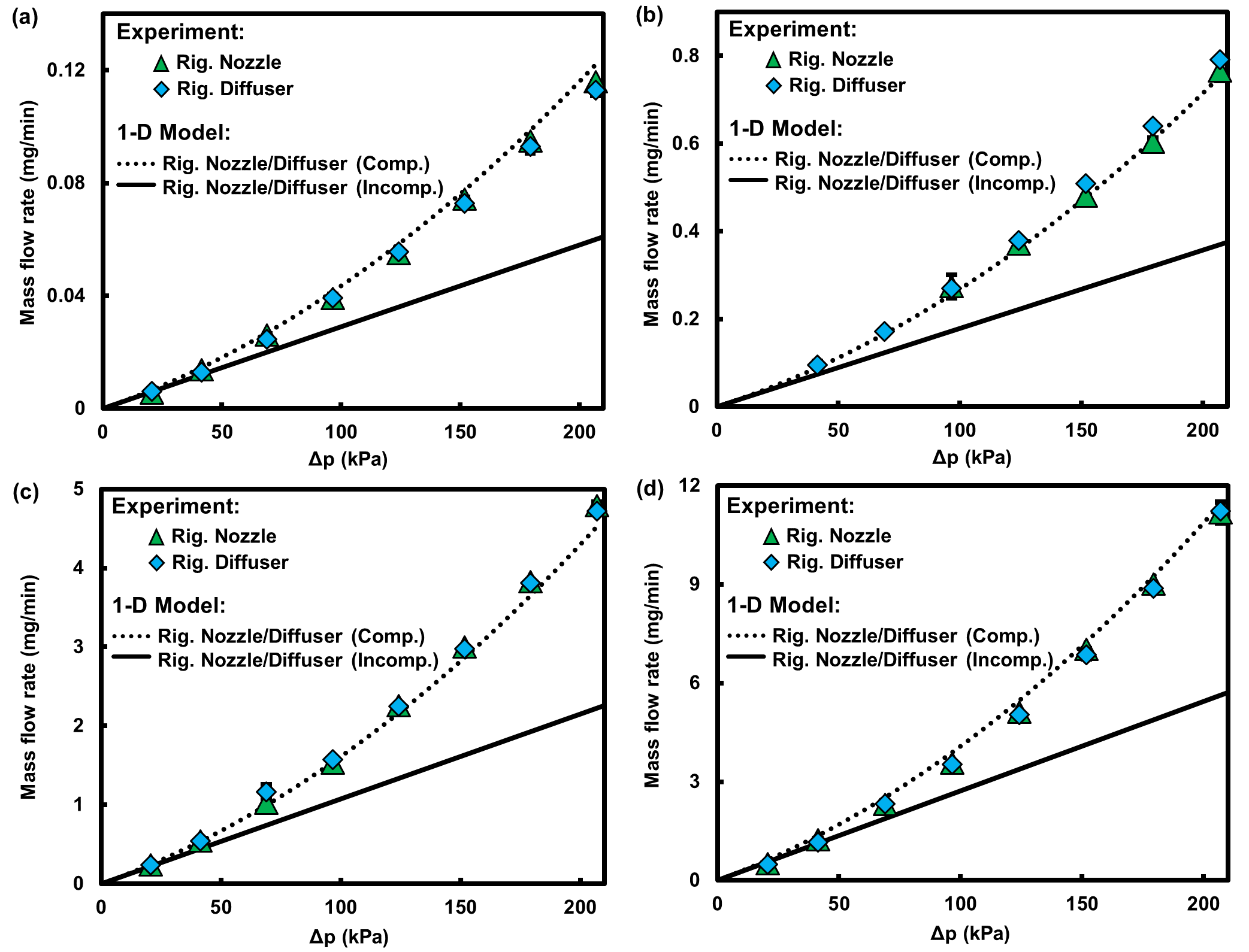}
	}
	\caption{Mass flow rate of air as a function of pressure difference across the rigid (Rig.) nozzle/diffuser microchannels with small and large widths of $1~mm$ and $2~mm$ and a half-angle of $1.25^\circ$ for various original heights of $H_0=$ (a) $2.6~\mu m$, (b) $4.6~\mu m$, (c) $8.0~\mu m$, and (d) $10.9~\mu m$.
	The modeling results with and without compressibility effects taken into account are shown by dotted and solid lines, respectively.
	}
	\label{Fig-Res-Flow-Rate-Rigid-Channels}
\end{figure*}

\begin{figure*}[!hbt]
\centering
	\subfloat{%
	\includegraphics[width=0.8\textwidth]{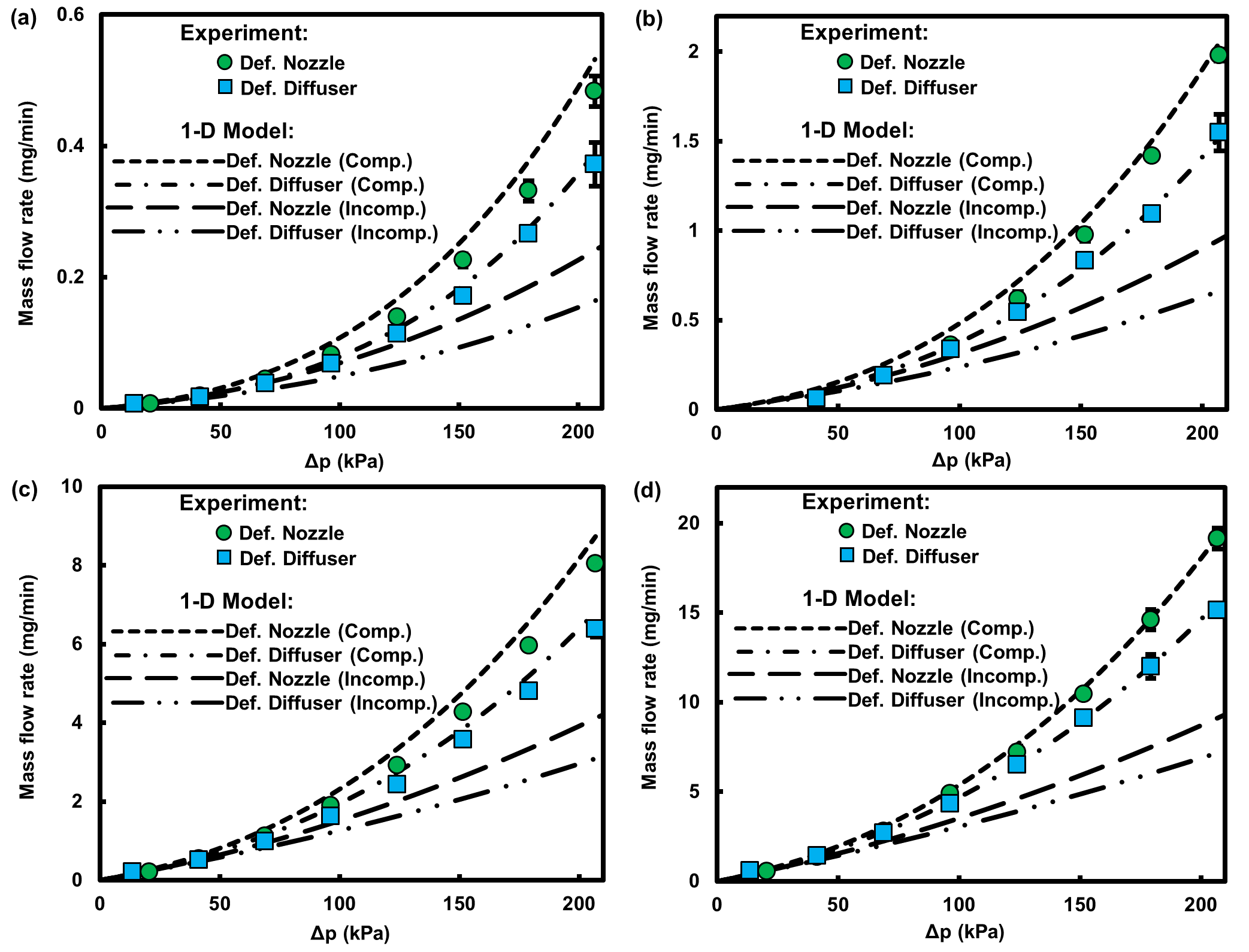}
	}
	\caption{Mass flow rate of air as a function of pressure difference across the deformable (Def.) nozzle/diffuser microchannels with small and large widths of $1~mm$ and $2~mm$ and a half-angle of $1.25^\circ$ for various original heights of $H_0=$ (a) $2.6~\mu m$, (b) $4.6~\mu m$, (c) $8.0~\mu m$, and (d) $10.9~\mu m$, together with modeling results for the air flow being considered as a compressible (Comp.) and incompressible (Incomp.) flow.
	}
	\label{Fig-Res-Flow-Rate-Deformable-Channels}
\end{figure*}

In our other work \cite{mehboudi_one-dimensional_2018}, we have already shown the importance of both fluid compressibility and microchannel deformability in studying the coupled fluid-solid mechanics, demonstrating that neglecting either of them under sufficiently large pressure differences can lead to erroneous results. 
To examine the compressibility effects on the characteristic behavior of the flow, particularly the rectification ratio, we have investigated air flow through the fabricated nozzle/diffuser devices.
The mass flow rates of the air flow through nozzle/diffuser microchannels with different original heights obtained from both modeling and experiments are shown in Figs.~\ref{Fig-Res-Flow-Rate-Rigid-Channels} and \ref{Fig-Res-Flow-Rate-Deformable-Channels} for rigid and deformable microchannels, respectively.
It is perceived from Fig.~\ref{Fig-Res-Flow-Rate-Rigid-Channels} that unlike an incompressible fluid flow through a rigid microchannel, the mass flow rate of the air flow increases nonlinearly with pressure difference.
Yet the rigid nozzle and diffuser exhibit the same characteristic behavior resulting in no flow rectification, {\it i.e.} the hydrodynamic resistance is not dependent on the fluid flow direction.
For the case of deformable microchannels, however, a nozzle delivers a larger mass flow rate under the same pressure difference, resulting in a fluid flow rectification; see Fig.~\ref{Fig-Res-Flow-Rate-Deformable-Channels}.

\begin{figure}[!hbt]
\centering
	\subfloat{%
	\includegraphics[width=0.5\textwidth]{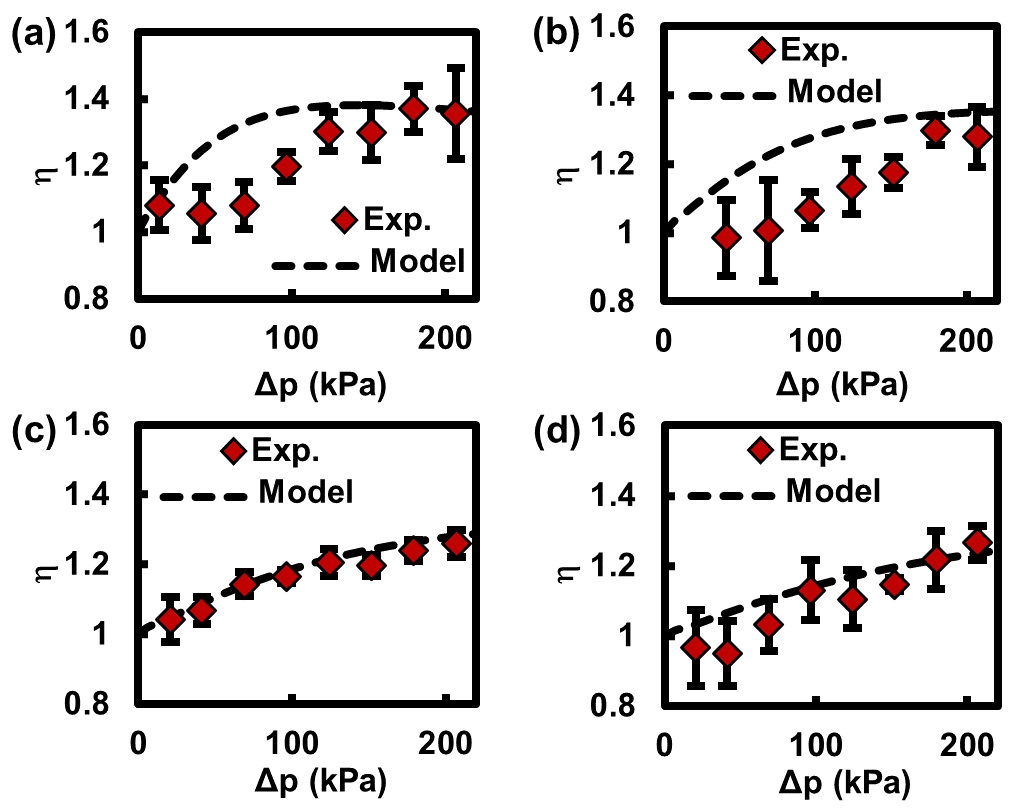}
	}
	\caption{Experimental and theoretical values of the air flow rectification ratio, {\it i.e.} $\eta=\dot{m}_\text{Nozzle}/\dot{m}_\text{Diffuser}$, as functions of pressure difference across the deformable nozzle/diffuser microchannels with small and large widths of $1~mm$ and $2~mm$ and a half-angle of $1.25^\circ$ for various original heights of $H_0=$ (a) $2.6~\mu m$, (b) $4.6~\mu m$, (c) $8.0~\mu m$, and (d) $10.9~\mu m$.
	}
	\label{Fig-Res-Rectification-Channels}
\end{figure}

The Knudsen number, {\it i.e.} $\text{Kn}\equiv\lambda/H=\sqrt{\pi/2R_\text{sgc}T_\text{atm}}(\mu/\rho H)$, where $\lambda$ denotes the mean free path of the gas molecules, has been calculated for different microchannels under various pressures.
For the microchannels studied in this work, the average Knudsen number $\text{Kn}_\text{avg}=(\text{Kn}_\text{i}+\text{Kn}_\text{o})/2$, where $\text{Kn}_\text{i}$ and $\text{Kn}_\text{o}$ refer to the Knudsen numbers at inlet and outlet, respectively, varies between $\sim$ 0.003 for highest pressure difference across the deepest channel and $\sim$ 0.03 for lowest pressure difference across the shallowest channel, which partially falls under the slip-flow regime category for few data points related to the low pressure differentials across the shallowest channel.
Although the Navier-Stokes equations are still applicable to this Kn range, the velocity slip on the walls can cause the mass flow rate to be underestimated.
The slip boundary condition can be implemented to Eq.~\ref{Eq-ND-Q} in order to derive a one-dimensional model suitable for the compressible Stokes flows under slip-flow regime.


\end{document}